\documentclass[useAMS,usenatbib]{mn2e}

\usepackage{graphicx}
\usepackage{txfonts}

\title[Planets in the Habitable Zone of binary stars with a gas giant]{Can there be additional rocky planets in the Habitable Zone of tight binary stars with a known gas giant?}
\author[B. Funk, E. Pilat-Lohinger and S. Eggl]{B. Funk$^{1}$\thanks{E-mail:
funk@astro.univie.ac.at; }, E. Pilat-Lohinger$^{1}$ and S. Eggl$^{2}$\\
$^{1}$Institute for Astronomy, University of Vienna, Vienna, Austria\\
$^{2}$IMCCE, Observatoire de Paris, Paris, France}
\begin{document}

\date{}

\pagerange{\pageref{firstpage}--\pageref{lastpage}} \pubyear{}

\maketitle

\label{firstpage}

\begin{abstract}
Locating planets in Habitable Zones (HZs) around other stars is a growing field in contemporary astronomy. 
Since a large percentage of all G-M stars in the solar neighborhood are expected to be part of binary or multiple stellar systems, 
investigations of whether habitable planets are likely to be discovered in such environments
are of prime interest to the scientific community. As current exoplanet statistics predicts that the chances are higher to find
new worlds in systems that are already known to have planets, 
we examine four known extrasolar planetary systems in tight binaries in order to determine their capacity to host additional habitable terrestrial planets.
Those systems are Gliese 86, $\gamma$ Cephei, HD 41004 and HD 196885. 
In the case of $\gamma$ Cephei, our results suggest that only the M dwarf companion could host additional potentially habitable worlds. 
Neither could we identify stable, potentially habitable regions around HD 196885 A. HD 196885 B can be considered a slightly more promising target in the search for Earth-twins.
Gliese 86 A turned out to be a very good candidate, assuming that the system's history has not been excessively violent. 
For HD 41004 we have identified admissible stable orbits for habitable planets, but those strongly depend on the parameters of the system. 
A more detailed investigation shows that for some initial conditions stable planetary motion is possible in the HZ of HD 41004 A. In spite of the massive companion HD41004 Bb we found that HD41004 B, too, could host additional habitable worlds.  
\end{abstract}


\section{Introduction}
The past decade has seen a staggering growth in number and diversity of known extrasolar planets. 
The discovery of planetary companions in and around binary stars systems is certainly a highlight in this respect \citep[e.g.][]{latham-1989,thorsett-1993,doyle-et-al-2011,welsh-et-al-2012,kostov-et-al-2014,howard-et-al-2014}.
We now know of more than 60 multiple star systems that harbor one or more extrasolar planets \citep{schneider-2011,rein-2014}.\footnote{http://www.openexoplanetcatalogue.com/, http://www.exoplanet.eu/}
Some of those systems, e.g. KIC 9632895 and Gliese 667 \citep{welsh-et-al-2014,anglada-et-al-2012} are even hosting planets in their respective Habitable Zones (HZ),
i.e. the regions where liquid water can exist on the surface of an Earth-like planet \citep[e.g.][]{kasting,kaltenegger,kopparapu}. There are also works, which showed that planet formation is possible in binary systems, even if the two stars have a quite small separation (e.g. \cite{haghighipour2007}, \cite{rafikov}). It is interesting to note that nearly one third of the currently known exoplanets in S-type\footnote{In this configuration the planet moves in an orbit around one of the stars.} configurations are part of multi-star multi-planet systems, i.e.
of systems where more than one planet is orbiting the same star (see e.g. \cite{rein-2014} or the Binary Catalogue of Exoplanets\footnote{http://www.univie.ac.at/adg/schwarz/multiple.html}).
Current statistics of the Kepler candidates exoplanet population furthermore suggest that about 46\% of all planets discovered so far reside in multiplanet systems \citep{burke-et-al-2014}.
Despite the fact that the planetary multiplicity fraction in multiple-star systems is not well constrained \citep{thorp-et-al-2014,ginski-et-al-2014,armstrong-et-al-2014},   
the high ratio of multiplanet systems in Kepler data 
may foster hopes to discover new planet candidates in multiple star associations that are already known to host a giant planet.
Such additional planets would have to be small or far out from their host star on long periodic orbits, otherwise they would have been detected by now \citep{eggl-et-al-2013b}. 
In this study we investigate whether it is possible for binary star systems with a known gas giant (GG) to host additional terrestrial planets in their respective HZs, since the latter are attractive observational targets. In particular we selected the following four binary systems with separation of approximately 20 au, since the strongest gravitational perturbations can be expected in such systems.

\begin{itemize}
\item {\bf $\gamma$ Cephei:}
The $\gamma$~Cephei binary system consists of a K1 III-IV star and a dwarf companion of spectral type M4. 
\citet{neuh} determined the orbital parameters of the binary system (as shown in Table~\ref{init}). Collecting RV measurements spanning many years \citet{endl} updated the orbital parameters and \citet{reffert} used Hipparcos data to search for the astrometric signature of planets and brown dwarfs. For the system $\gamma$~Cephei the signature of the substellar companion 
was strong enough to give reasonable constraints on inclination, ascending node and mass. They claim that $\gamma$~Cephei~b is most likely a brown dwarf. 
\item {\bf Gliese 86:}
\citet{queloz} discovered a 4 $M_{Jup}$ planet around the K1 V star Gliese 86. One year later \citet{els} announced a substellar companion with a mass of approximately 50 $M_{Jup}$ in a distance of 18.75 au. Later \citet{mugrauer} confirmed that the secondary is a White Dwarf (WD) and not a brown dwarf. With the help of new observations using the Hubble Space Telescope \citet{farihi} could proof that Gliese 86 B is a WD and could give modified data for the system.
\item {\bf HD 196885:}
The system HD 196885 consists of an F8 V star with 1.3 $M_{Sun}$, a M1 V companion and a planet orbiting the primary (see e.g. \citet{chauvin06}, \citet{chauvin07}, \citet{correia} or \citet{fischer}). \citet{chauvin} considered astrometric data as well as all data from past RV surveys (CORAVEL, ELODIE, CORALIE and Lick, over more than 26 years) to constrain the physical and orbital properties of the HD 196885 system and could find a unique best solution, which is summarized in Table~\ref{init}. Furthermore \citet{chauvin} investigated the stability of the system and showed that the system is more stable in a high mutual inclination configuration.
\item {\bf HD 41004:}
The system HD 41004 comprises a K1 V primary (HD 41004 A) and an M2 V secondary (HD 41004 B) at a distance of 20~au. \citet{santos02} discovered a brown dwarf orbiting HD 41004 B and later \citet{zucker} found a GG around HD 41004 A.
\end{itemize}
In the following we investigate, whether dynamical stability allows for an additional terrestrial planet in the HZs of the four considered systems.

\begin{table*}
 \centering
  \caption{Keplerian orbital parameters and errors (if given) of the four known tight binary systems, hosting an extrasolar planet. The orbital elements of component B are given with respect to component A and the elements of the planets are given with respect to the respective central star. Given are the Name, semi-major axis ($a$), eccentricity ($e$), argument of perihelion ($\omega$), Longitude of the ascending node ($\Omega$), mass, spectral type and age.}
  \begin{tabular}{l|llllllll|}
  \hline
   Name & $a$ [au] & $e$ & $\omega$ [deg] &  $\Omega$ [deg] & mass & Sp.Type & Age [Gyr] & Ref.\\
 \hline
$\gamma$ Cephei A & - & - & - & - & 1.4 $M_{Sun}$ & K1 III-IV & 6.6 & \citet{neuh}\\
 &  &  &  &  &  $\pm$ 0.12 & & \\
$\gamma$ Cephei B & 20.18 & 0.4112 & 162.0 & 18.04 & 0.409 $M_{Sun}$ & M4 && \citet{neuh}\\
& $\pm$ 0.66 & $\pm$ 0.0063 & $\pm$ 0.40 & $\pm$ 0.98 & $\pm$ 0.018 & & &\\
$\gamma$ Cephei Ab & 2.05 & 0.049 & 94.6 & 37.5 & 1.85 $M_{Jup}$ & - & &\citet{endl},\\
&  & &  & &&&  &\citet{reffert}\\
& $\pm$ 0.06 & $\pm$ 0.034 & $\pm$ 34.6 & & $\pm$ 0.06 &  & &\\
\hline
Gliese 86 A & - & - & - & - &  0.8 $M_{Sun}$ & K0 V & 2.03 & \citet{farihi}\\
Gliese 86 B & 27.8 - 69.8 & 0.0 - 0.61 & - & - & 0.59 $M_{Sun}$ & DQ6 && \citet{farihi}\\
&  & &  &  & $\pm$ 0.01 & & &\\
Gliese 86 Ab & 0.11 & 0.046 & 270.0 & - & 4.0 $M_{Jup}$ & - & &\citet{queloz}\\
&  & $\pm$ 0.004 & $\pm$ 4 &  &  &  & &\\
\hline
HD 196885 A & - & - & - & - & 1.3 $M_{Sun}$ & F8 V & 2.0 & \citet{chauvin}\\
 &  &  &  &  &  &  & $\pm$ 0.5 & \\
HD 196885 B & 21 & 0.42 & -118.1 & 79.8 & 0.45 $M_{Sun}$ & M1 V & &\citet{chauvin}\\
& $\pm$ 0.86 & $\pm$ 0.03 & $\pm$ 3.1 & $\pm$ 0.1 & $\pm$ 0.1 &  &&\\
HD 196885 Ab & 2.6 & 0.48 & 93.2  & - & 2.98 $M_{Jup}$ & - && \citet{chauvin}\\
& $\pm$ 0.1 & $\pm$ 0.02 & $\pm$ 3.0 &  & $\pm$ 0.05 &  & &\\
\hline
HD 41004 A & - & - & - & - & 0.7 $M_{Sun}$ & K1 V & 1.64  & \citet{santos02}\\
HD 41004 B & 20 & 0.4 & - & - & 0.42 $M_{Sun}$ & M2 V & 1.56 & \citet{roell}, \citet{zucker}\\
&  &  &  &  &  &  & $\pm$ 0.8 & \\
HD 41004 Ab & 1.6 & 0.39 & 97 & - & 2.54 $M_{Jup}$ & - & & \citet{roell}, \citet{zucker}\\
 & & $\pm$ 0.17 & $\pm$ 31 &  & $\pm$ 0.74 &  && \\
HD 41004 Bb & 0.0177 & 0.081 & 178.5 & - & 18.37 $M_{Jup}$ & - && \citet{zucker}, \citet{santos02}\\
& & $\pm$ 0.012 & $\pm$ 7.8 &  & $\pm$ 0.22 &  &&\\
\hline
\end{tabular}
\label{init}
\end{table*}

\section{The Habitable Zones}
The definition of the HZ is a highly interdisciplinary topic. Important parameters are not only the mass and atmosphere of the potentially habitable planet \citep{kopparapu}, but also properties of the host star like activity and radiation (e.g. \citep{lammer}) and last but not least the dynamical stability of all involved bodies play a significant role. 
The evolution of orbital parameters can also be a determining factor. Large variations in a planet's orbital eccentricity, for instance, 
might change the conditions for habitability (e.g. \citep{willi}).
Considerations of dynamical stability are of special interest for multiple star planet systems due to potentially large gravitational perturbations of the planet's orbit \citep{jaime-2014}. 
\cite{eggl} investigated the influence of stellar companions with different spectral types on the insolation a terrestrial planet receives orbiting a Sun-like primary. 
They used a time-independent analytical estimate to calculate the borders of the HZ and found out that there is a strong dependence of permanent habitability on the binary's eccentricity. With four bodies in a system, special attention has to be given to the assessment of habitability. The following section contains a brief discussion on the 
related issues.

\subsection{Calculating the borders of the HZs}
\label{hz}
\begin{figure*}
\includegraphics[width=5.9cm, angle=270]{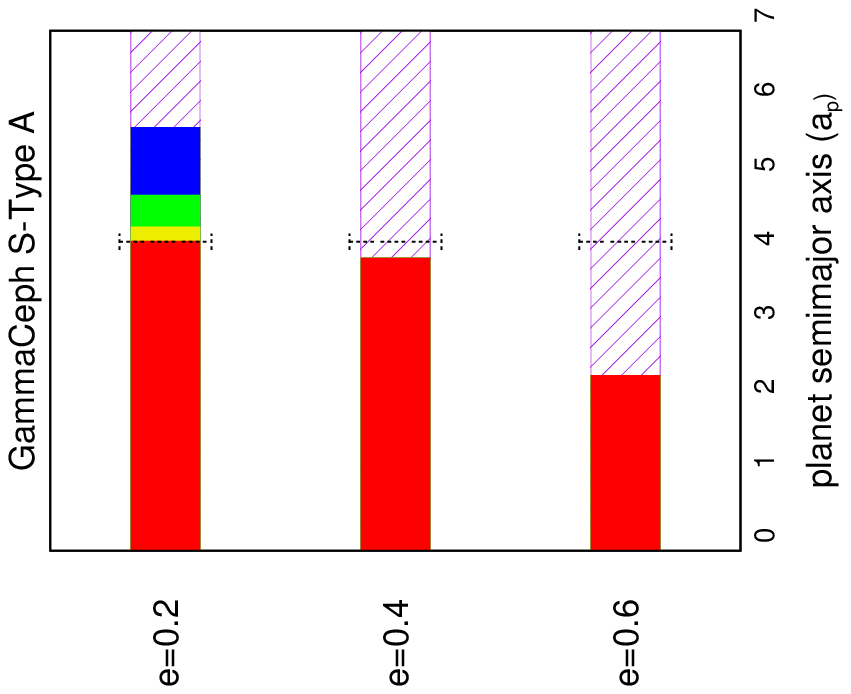}
\includegraphics[width=5.9cm, angle=270]{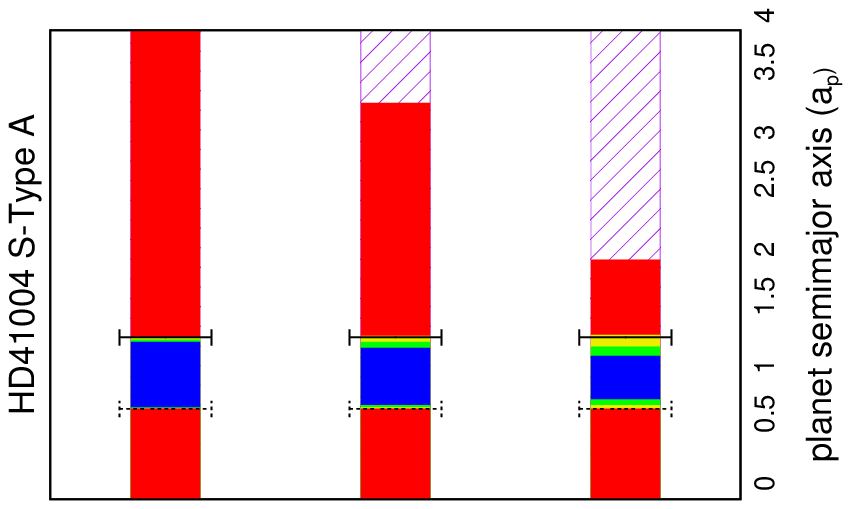}
\includegraphics[width=5.9cm, angle=270]{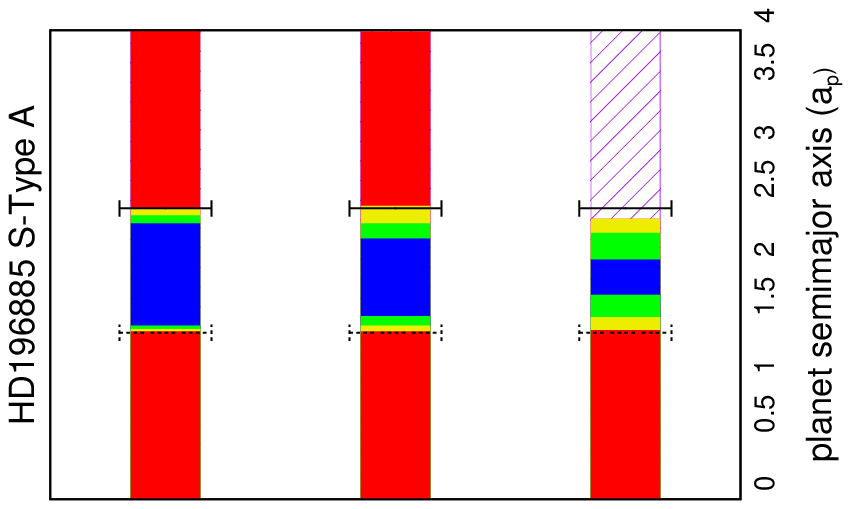}
\includegraphics[width=5.9cm, angle=270]{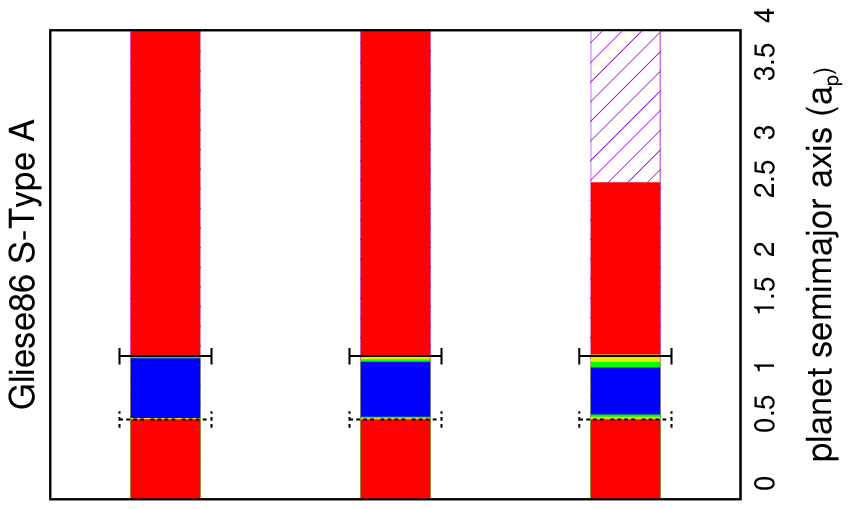}
\caption{Circumprimary HZs for the systems (from left to right) $\gamma$ Ceph, HD 41004 , HD 196885 and Gliese 86, where blue gives the PHZ, green the EHZ and yellow the AHZ. Red regions are not habitable, but dynamically stable. Shaded regions are dynamically unstable. The black lines gives the extent of the HZ, when ignoring the secondary star. See online version for color figures. \label{fig1}}
\label{hzhdhd}
\end{figure*}

In this section we will shortly summarize how to determine the HZ for the four investigated binary star systems. Several approaches can be found in literature how to calculate the borders of the circumstellar HZ in binary star systems (see e.g. \cite{cuntz}, \cite{kaltenegger2013}). We choose the method presented in \cite{eggl}, because it combines the radiative and the dynamical effects.
\cite{eggl} investigated different binary-planet configurations and examined the implications of stellar companions with different spectral types on the insolation a 
terrestrial planet receives orbiting a Sun-like primary. The time-independent analytical estimates that were presented in \cite{eggl}
are based on three types of HZs:
\begin{itemize}
\item {\bf Permanently Habitable Zone (PHZ)}: The PHZ is the region where a planet always stays within the insolation limits of the corresponding HZ. 
\item {\bf Extended Habitable Zone (EHZ)}: In contrast to the PHZ parts of the planetary orbit lie outside the HZ. Yet, the binary planet configuration is still considered to be habitable when most of its orbit remains inside the HZ boundaries.
\item {\bf Averaged Habitable Zone (AHZ)}: Following the argument by \cite{willi}, this category encompasses all configurations which allow for the planet's time-averaged effective insolation to be within the limits of the HZ.
\end{itemize}
\cite{eggl} found that the AHZ is almost independent of the binary's eccentricity and coincides well with the HZ defined by \cite{kasting}, while the PHZ and EHZ shrink drastically with higher binary eccentricities. The gravitational influence of the second star injects eccentricity into the  planet's orbit, which in turn leads to considerable changes in planetary insolation due to closer encounters with the host-star.\\
The HZs of $\gamma$ Cephei, HD 41004, HD 196885 and Gliese 86 were determined using the analytical method presented in \cite{eggl} with updated effective insolation values \citep{kopparapu}. The results are shown in Figures~\ref{hzhdhd}.\\

\begin{figure*}
\includegraphics[width=5.9cm, angle=270]{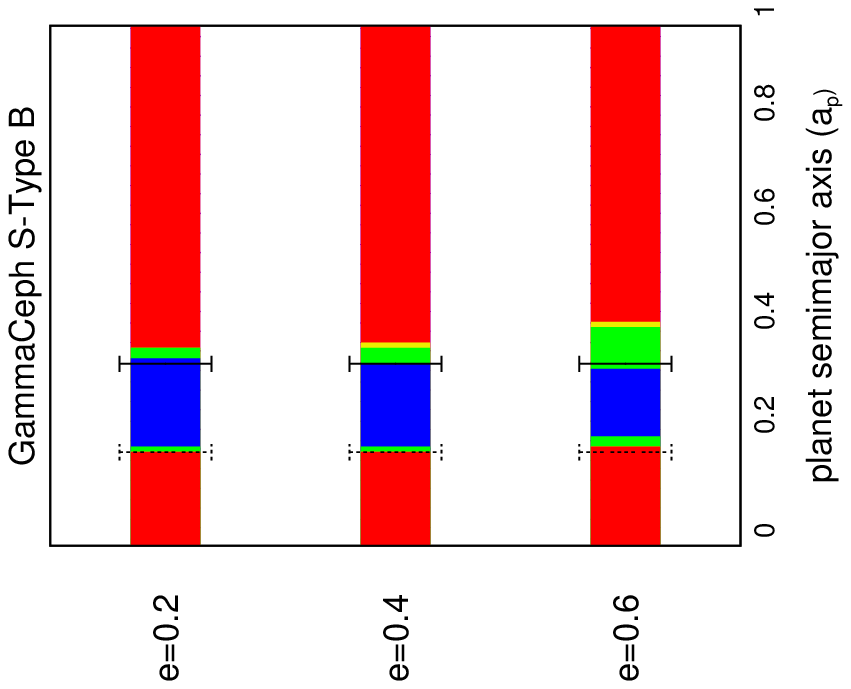}
\includegraphics[width=5.9cm, angle=270]{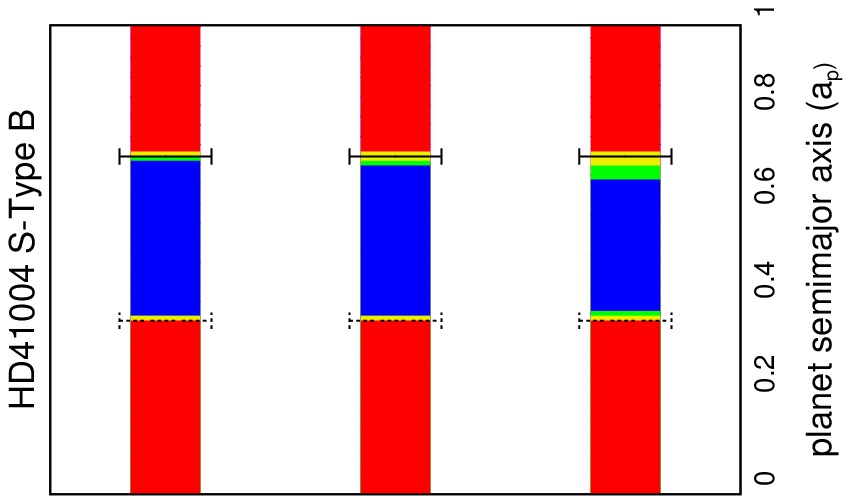}
\includegraphics[width=5.9cm, angle=270]{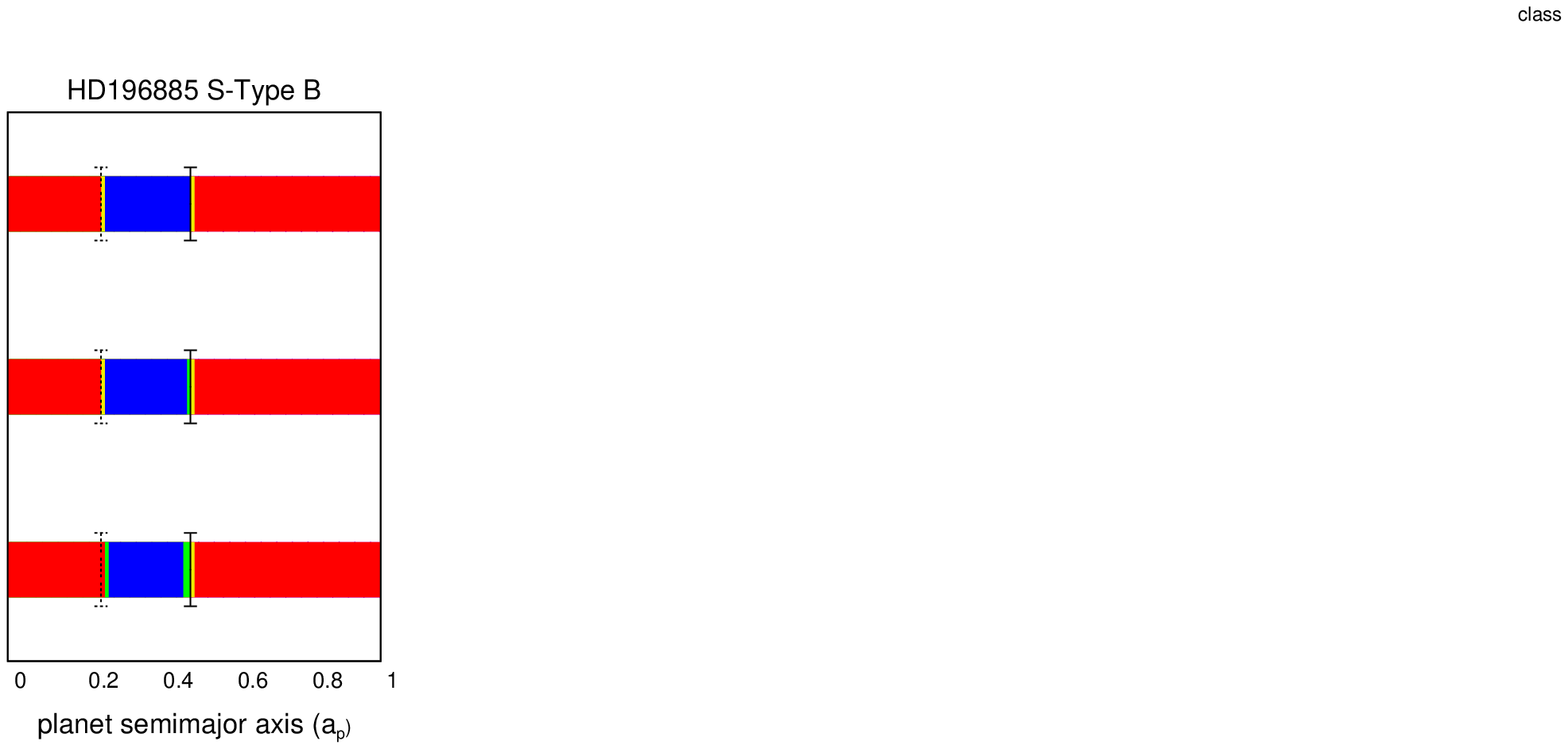}
\includegraphics[width=5.9cm, angle=270]{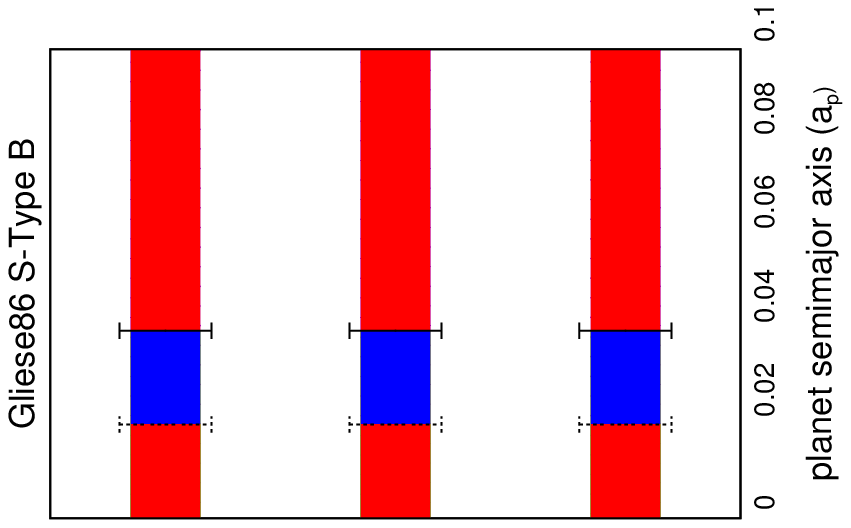}
\caption{Same as Figrue \ref{fig1}, only for HZs around the less massive component (B) of the binary star. From left to right: $\gamma$ Ceph, HD 41004 , HD 196885 and Gliese 86. Note the difference in the scale of the planetary semimajor axis for Gliese 86. See online version for color figures.}
\label{hzhdhd2}
\end{figure*}

Blue regions denote the extent of the PHZ, green regions correspond to EHZs and yellow represents AHZs. 
Red regions are not habitable, but dynamically stable, while the shaded regions are dynamically unstable. 
The black lines corresponds to the classical HZ where the second star is completely neglected. The initial conditions for the binaries are summarized in Table~\ref{init}. 
Due to the large orbital periods of S-type binary stars, the eccentricity of the binary systems is often poorly constrained. 
Hence, results for three different eccentricity values of the binary's orbit ($e_{Bin}$ = 0.2, 0.4 and 0.6) are provided.\\
The borders of the PHZ, EHZ and AHZ are summarized in Table~\ref{hztab}.
For all four systems one can see that the PHZ shrinks with higher eccentricities. A comparison of our work with that of \cite{kaltenegger2013} can be done for the system HD~196885, which was investigated in both studies. \cite{kaltenegger2013} determined the following borders for the HZs of components A and B\footnote{Considering the maximum flux of the companion, which corresponds to the closest approach of the stars and therefore, to the strongest influence on the HZ. Since no dynamical perturbations are considered in \cite{kaltenegger2013} the HZ borders for the minimum and maximum flux values are almost the same (deviations between 0.002 au and 0.008 au).}: ${HZ_A}$: 1.454 au - 2.477 au; ${HZ_B}$: 0.260 au - 0.491 au. Comparing our results, we find that the PHZ lies within the borders of \cite{kaltenegger2013}, which is due to the consideration of the perturbing influence of the secondary. The EHZ allows eccentric planetary orbits, that may leave the HZ for a short time, which leads to quite similar results as in \cite{kaltenegger2013}.\\
Please note that the analytical method used to determine the presented HZ borders assumes that the luminosity of the stars does not evolve significantly. For stars on the main sequence this is true for billions of years. For post main sequence stars, the timescale on which those estimates remain valid depends on the speed at which the stellar luminosity evolves. Furthermore, the analytic solutions are based on eccentricity estimates that do not consider the known GG. They merely serve as initial guidelines to determine search spaces for additional, potentially habitable terrestrial planets. If the GG influences the eccentricity evolution of additional planets significantly, a separate assessment of their habitability is required, see section \ref{sec:hd41004}.

\begin{table*}
 \centering
  \caption{Values of the AHZ, EHZ and PHZ (in au) for the systems $\gamma$ Cephei, HD 41004, HD 196885 and Gliese 86. 
  The HZ borders are provided for the primary (A) and the secondary star (B).
  Since the eccentricity of some of the binaries is not well constrained we give three values for each system corresponding to different 
  binary star eccentricities: $e_{Bin} =$ 0.2, $e_{Bin} =$ 0.4, and $e_{Bin} =$ 0.6.}
  \begin{tabular}{l|llllll|}
  \hline
   Name & inner AHZ & inner EHZ & inner PHZ & outer PHZ & outer EHZ & outer AHZ\\
\hline
$\gamma$ Cephei A ($e_{Bin} =$ 0.2) & 4.17 &          4.36 &          4.79 &          5.70 &          5.70 &          5.70\\
$\gamma$ Cephei A ($e_{Bin} =$ 0.4) & -& -& -& -& -&- \\
$\gamma$ Cephei A ($e_{Bin} =$ 0.6) & -& -& -& -& -&- \\\hline
$\gamma$ Cephei B ($e_{Bin} =$ 0.2) &  0.183 &         0.183 &         0.187 &         0.363 &         0.377 &         0.379\\
$\gamma$ Cephei B ($e_{Bin} =$ 0.4) & 0.185 &         0.185 &         0.193 &         0.355 &         0.385 &         0.389\\
$\gamma$ Cephei B ($e_{Bin} =$ 0.6) & 0.189 &         0.191 &         0.215 &         0.345 &         0.421 &         0.431\\
\hline
HD 41004 A ($e_{Bin} =$ 0.2) & 0.77 &          0.77 &          0.78 &          1.34 &          1.36 &          1.38\\
HD 41004 A ($e_{Bin} =$ 0.4) & 0.77 &          0.78 &          0.80 &          1.29 &          1.34 &          1.39\\
HD 41004 A ($e_{Bin} =$ 0.6) & 0.77 &          0.80 &          0.85 &          1.22 &          1.30 &          1.40\\\hline
HD 41004 B ($e_{Bin} =$ 0.2) & 0.374 &         0.376 &         0.378 &         0.712 &         0.720 &         0.726\\
HD 41004 B ($e_{Bin} =$ 0.4) &  0.374 &         0.378 &         0.382 &         0.696 &         0.712 &         0.726\\
HD 41004 B ($e_{Bin} =$ 0.6) &  0.374 &         0.382 &         0.392 &         0.674 &         0.704 &         0.730\\
\hline
HD 196885 A ($e_{Bin} =$ 0.2) &  1.43 &          1.45 &          1.48 &          2.35 &          2.42 &          2.48\\
HD 196885 A ($e_{Bin} =$ 0.4) & 1.43 &          1.48 &          1.56 &          2.22 &          2.35 &          2.50\\
HD 196885 A ($e_{Bin} =$ 0.6) & 1.44 &          1.55 &          1.74 &          2.04 &          2.27 &          2.39\\\hline
HD 196885 B ($e_{Bin} =$ 0.2) & 0.255 &         0.257 &         0.257 &         0.489 &         0.493 &         0.497\\
HD 196885 B ($e_{Bin} =$ 0.4) & 0.255 &         0.257 &         0.261 &         0.481 &         0.491 &         0.497\\
HD 196885 B ($e_{Bin} =$ 0.6) & 0.257 &         0.259 &         0.267 &         0.471 &         0.489 &         0.501\\
\hline
Gliese 86 A ($e_{Bin} =$ 0.2) &  0.68 &          0.69 &          0.69 &          1.20 &          1.21 &          1.22\\
Gliese 86 A ($e_{Bin} =$ 0.4) & 0.68 &          0.69 &          0.70 &          1.17 &          1.19 &          1.22\\
Gliese 86 A ($e_{Bin} =$ 0.6) & 0.68 &          0.70 &          0.72 &          1.12 &          1.17 &          1.23\\\hline
Gliese 86 B ($e_{Bin} =$ 0.2) & 0.023 &         0.023 &         0.023 &         0.037 &         0.037 &         0.037\\
Gliese 86 B ($e_{Bin} =$ 0.4) & 0.023 &         0.023 &         0.023 &         0.037 &         0.037 &         0.037\\
Gliese 86 B ($e_{Bin} =$ 0.6) & 0.023 &         0.023 &         0.023 &         0.037 &         0.037 &         0.037\\\hline
\hline
\end{tabular}
\label{hztab}
\end{table*}

\subsection{Stability of the HZs}
\label{stab}
In order to ensure the dynamical stability of an additional terrestrial planet in the HZ of the four investigated systems we studied the dynamical evolution of 
test-planets within the borders of the previous calculated region of the HZ (see Table~\ref{hztab}) for each system. 
As a model we used the elliptic restricted four-body problem, consisting of the two stars, the known GG and mass-less TP's, which is a good approximation for terrestrial planets. In the case of HD 41004 the masses of component B and of the nearby brown dwarf were added and put to there common barycenter, since test computations showed that the brown dwarf has no perturbing influence on the region of the HZ (see section 6.3). In all cases, the motion of the planets and the binary star were treated self consistently  in the framework of Newtonian mechanics.
All the celestial bodies involved were regarded as point masses. Test computations have shown that a time of $10^{6}$ years is sufficient to capture the most relevant dynamical effects. For all integrations we used the Lie-Series Integration Method (\citet{13}, \citet{20}, \citet{34}) 
which is also capable of dealing with large eccentricities and close encounters between bodies. To analyze the data we used the maximum eccentricity (ME) that
the test planets (TP) orbit reached during the integration time. If the TP's orbit becomes parabolic or hyperbolic (ME $\ge$ 1) with respect to it's central star the orbit is considered to be unstable. 
A further advantage of using ME for the investigations of TP's in the HZ is that information on the extent of the PHZ becomes directly available.\\
The orbital parameters of the investigated four systems are summarized in Table~\ref{init}. The mass-less TP's were started equally distributed ($\Delta a = 0.01$ au) within the borders of the HZ. All other orbital elements of the test planets ($e$, $i$, $\omega$, $\Omega$ and $M$) were initially set to zero.\\

\begin{figure}
\includegraphics[width=6.6cm]{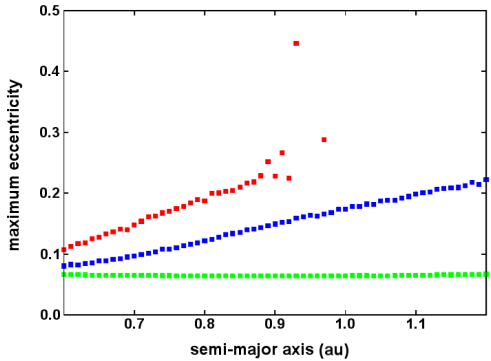}
\caption{Maximum eccentricities for three different eccentricities of Gliese 86 B (green: $e_{Bin}$ = 0.0, blue: $e_{Bin}$ = 0.7, red: $e_{Bin}$ = 0.8). On the $x$-axis we show the semi-major axis, covering the HZ. The $y$-axis gives the maximum eccentricity values. See online version for color figures.}
\label{gliesemaxe}
\end{figure}

\section{$\gamma$ Cephei}

The $\gamma$ Cephei system has already been investigated in detail concerning the dynamical stability in its circumprimary HZ  (see e.g. \citet{dvorak2} or \citet{hagh1}). However, the contribution of the secondary to the extent of the HZ was completely neglected in those studies. The numerical study by \cite{dvorak2} examined the area between 0.5 and 1.85 au concerning stability. The stability maps showed a stable region between 0.5 and 1.2 au and a chaotic area for lager semi-major axes of the test-planet. Within the stable zone an arched chaotic band was found which
results from the combined perturbation of the giant planet and the secondary star (see e.g. \citet{pilat}). The authors studied also the area around 1 au in detail using several masses (up to 90 Earth masses) for the small planet moving in this area. Even if they talked about habitability when discussing the results it is well known that the studied region does not correspond to the HZ of $\gamma$ Cephei. This was also criticized in the study by \cite{hagh1} where the area between 0.3 and 4 au was studied concerning long-term stability and habitability. Taking into account the stellar type Haghighipour defined the HZ as the area between 3.05 and 3.6 au for which he could not get long-term stable planetary orbits.\\
Updates for the HZ around $\gamma$~Cephei A show that the habitable limits are shifted beyond the orbital stability limits of the current system ($e_{Bin}=0.4112$). 
Only stable orbits in the HZ around the secondary are possible. As no planet is known to orbit $\gamma$~Cephei B the HZ borders given in Table 1 can
serve as an observational guideline provided one assumes that stable circulation patterns have formed on the additional terrestrial planet \citep{wang-et-al-2014} and 
tidal heating remains negligible.

\section{HD 196885}
For the system HD 196885 our results provide constraints on the possible eccentricity of the binary. 
Since we know a GG at 2.6 au the given eccentricity of the binary ($e$ = 0.42) is quite close to the system's stability limit.
Comparing Tables~\ref{init} and \ref{hztab} one can see that the known planet in the systems HD 196885 and HD 41004 orbits close to or even inside the HZ. 
For HD 196885 the GG is located in the middle of the HZ. Given its high eccentricity 
of $e$ = 0.48 it will cause dynamical instability for any planet in this region. Test computations showed that even 
if the known planet had no eccentricity ($e$ = 0) no additional stable orbits would be possible.
Similar to $\gamma$~Cephei, the only possibility for HD 196885 to host additional potentially habitable planets is to have them orbit the M-dwarf companion. 

\section{Gliese 86}
A detailed dynamical study on Gliese 86 has already been performed by \cite{lohinger}. They found that the GG does not dynamically influence the region of the HZ. 
Since the orbital parameters of Gliese 86 as well as the method of determining the HZ have changed we shall check their results for robustness. 
To this end, we compared the analytically obtained HZ (without GG, Figure~\ref{hzhdhd}) with numerical 
integrations of the whole system (including the GG). Since the binary star's orbit is not well constrained, we integrated this system for different eccentricities. Figure~\ref{gliesemaxe} shows the results for $e$ = 0.0, 0.7 and 0.8.
The $x$-axis the semi-major axis, covering the best-case HZ ($e_{Bin}=0$, see Table~\ref{hztab}). 
The $y$-axis gives the maximum eccentricity value. It turns out that for binary eccentricities up to $e$ = 0.7 the TP's maximum eccentricity reaches values up to 0.2 only. 
Thus we can conclude that stable habitable planets are possible in the Gliese 86 system even if the binary star's orbit is highly elliptic. 
Only for binary eccentricities higher than $e$ = 0.7 the secondary's influence on the test planets in the region of the HZ becomes critical. 
Comparing these results with the analytically obtained HZ borders (see Table~\ref{hztab} and Figure~\ref{hzhdhd}) one can see that similar results can be obtained for the test planets eccentricity, which means that the known GG has no significant influence on additional planets in the HZ. Therefore, the analytically derived borders of the HZ remain valid even for the whole system of Gliese 86. 
 
Nevertheless, we are faced with two problems when talking about habitable planets in Gliese 86.
The first issue is related to the fact that Gliese 86 B is a WD. This suggests a cataclysmic phase in the evolution of the Gliese 86 system
with strong variations in stellar insolation and mass loss. Whether an Earth-like planet could retain any sort of habitability during such events or can be captured afterward is discussed in e.g. \cite{agol}, \cite{barnes}, \cite{veras}. 
The second issue is the small distance of the known GG to its host star. This could imply that migration processes occurred either
during the system's formation, or during the cataclysmic phase.
If the GG formed beyond the Gliese 86 A's snowline and migrated inward, it had to cross the HZ. This would have destabilized any possible terrestrial planet in the circumprimary HZ.

However, two scenarios are thinkable that would still allow for the existence of potentially habitable worlds around Gliese 86 A:
a) the planet was built after the migration of the GG \citep[e.g.][]{raymond}, or b) the planet formed outside of the GG, was trapped in a mean motion resonance 
and migrated inward together with the GG. Such a scenario would allow for a large primordial water content of the rocky planet. 

While it is clear that Gliese86 B has gone through a cataclysmic phase, it is intriguing to note that a 
circumstellar habitable planet around the B component seems possible. Although a WD, Gliese 86 B does have a circumstellar region that would allow for liquid water
on an Earth-like planet's surface. This region is very small and very close to the star, as can be seen from Figure \ref{hzhdhd2}.
However, tidal forces do not yet outweigh the self gravity of Earth-like planets at this distance, since
the fraction of accelerations due to the planet's self gravity $a_{SG}$ and tides $a_T$ acting on a particle at the
top of the atmosphere 100km above the Earth's surface (K\'arm\'an line) is still much larger than unity:
\begin{equation}
\frac{a_{SG}}{a_{T}}\simeq \frac{m_p r^3}{2 m_\star R_p^3} \approx 40 
\end{equation}
for an Earthlike planet orbiting Gliese86 (B) in the HZ.
Here, $m_p$ and $R_p$ denote the planet's mass and radius to the top of the atmosphere, respectively.
Furthermore, $m_\star$ is the stellar mass and $r$ the planet's orbital distance from the host star. 
Hence, if an Earth-like planet had survived the cataclysmic event, or if it had formed later on, it could retain its atmosphere.
If it orbits in the HZ it could be potentially habitable - from the point of view
of insolation. A tidally locked rotation state and non negligible tidal heating would be likely in such environments. 
Such factors require a more detailed analysis that lies beyond the scope of this article.

\section{HD 41004}
\label{reshz41004}
\subsection{Dynamical Study}

\begin{figure}
\includegraphics[width=6.6cm]{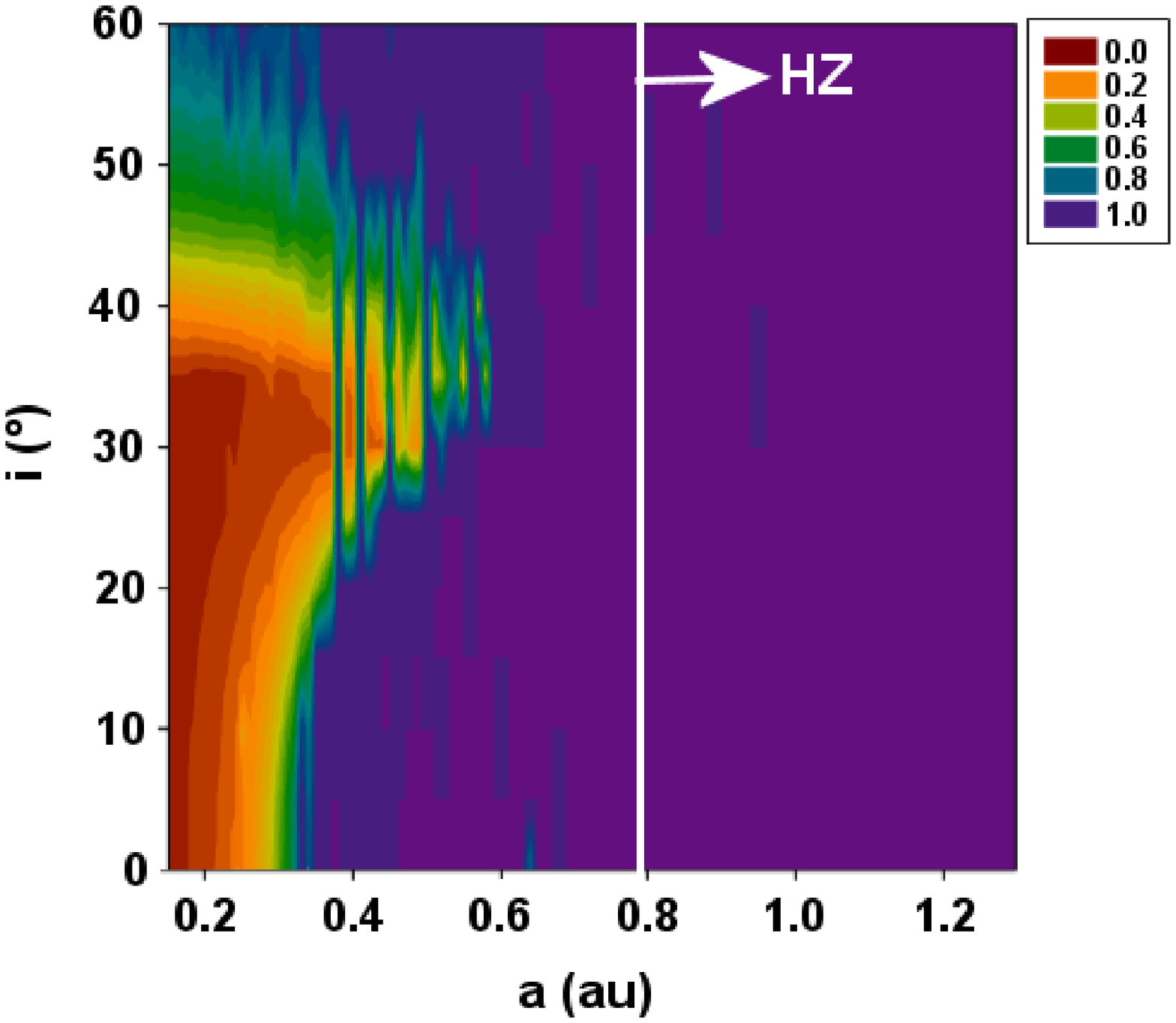}
\includegraphics[width=6.6cm]{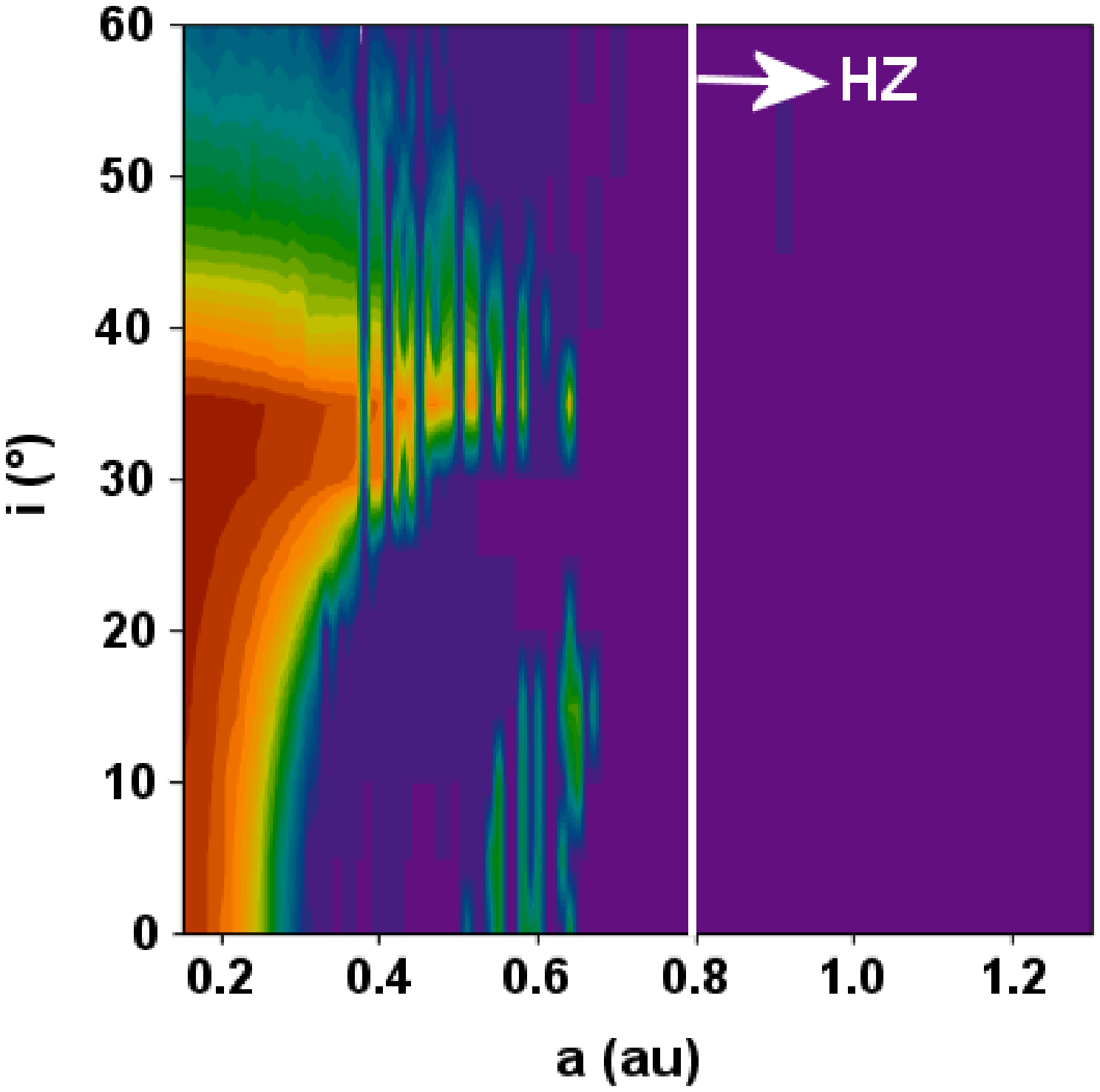}
\includegraphics[width=6.6cm]{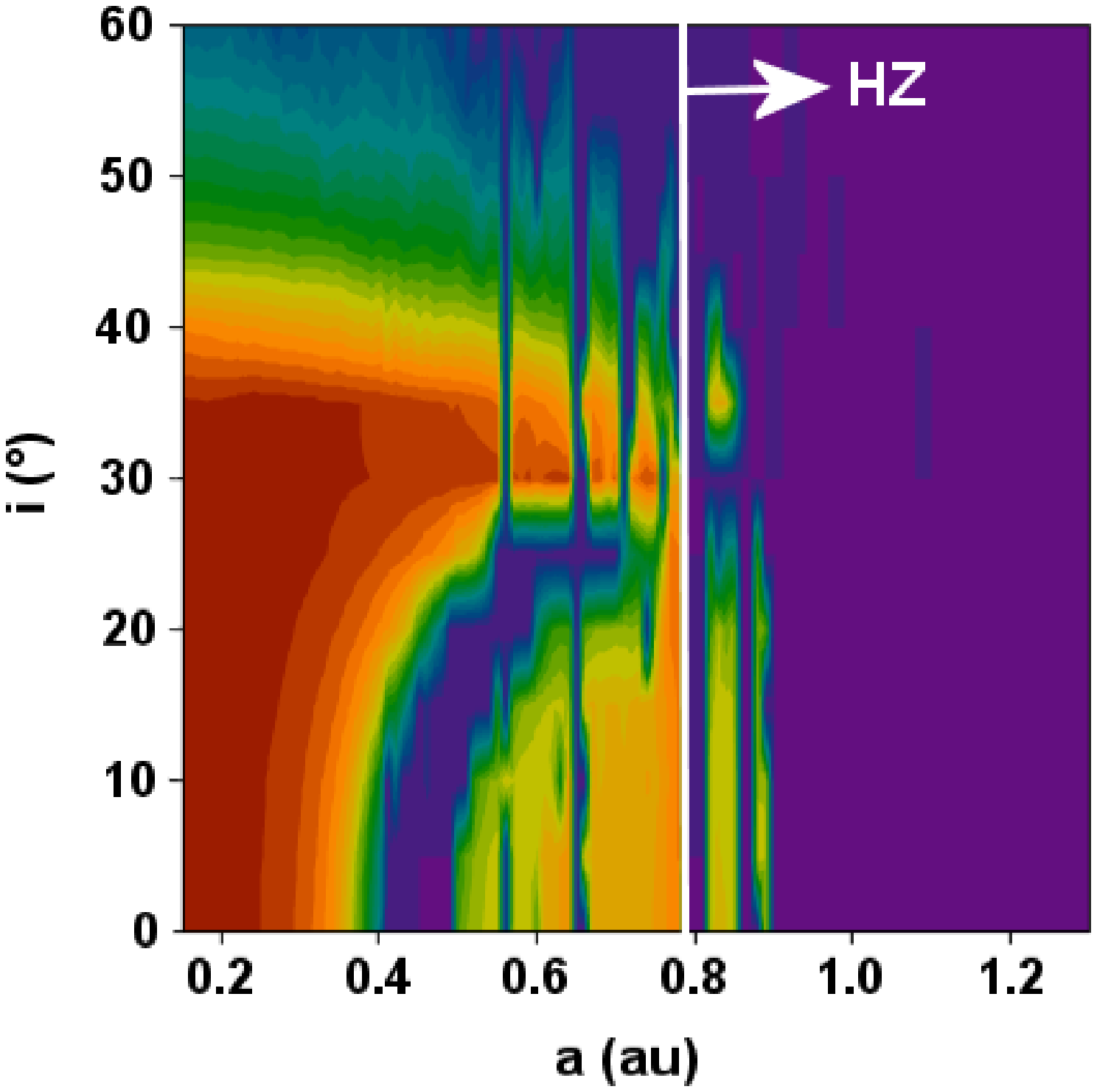}
\caption{Semi-major axis of the test planets versus their inclination. Upper graph: given orbital parameters for the system HD 41004; middle graph: like upper graph, but for $e_{Bin}$ = 0.2; lower graph: like middle graph, but for $e_{GG}$ = 0.2. The color code corresponds to the maximum eccentricity of the test planets, where violet shows unstable motion. The white vertical line shows the inner border of the PHZ. See online version for color figures.}
\label{e_hd41004}
\end{figure}

\begin{figure}
\includegraphics[width=6.6cm]{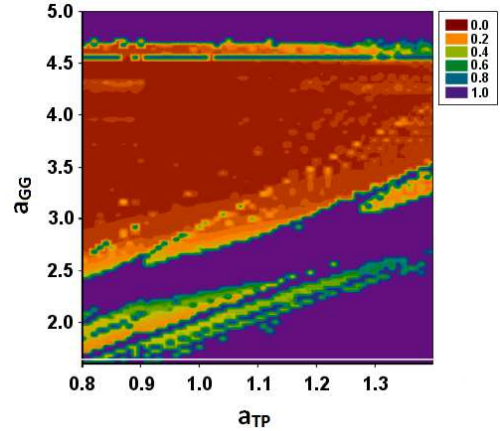}
\caption{Semi-major axis of the test planets versus semi-major axis of the gas giant for an eccentricity of the planet of $e_{GG} = 0.2$ and an eccentricity of the binary of $e_{Bin} = 0.2$. All other orbital elements are as shown in Table~\ref{init}; the test planets move on a planar orbit. The color code corresponds to the maximum eccentricity of the test planets, where violet shows unstable motion. The white line indicates the current position of HD 41004 A b. See online version for color figures.}
\label{atpaggi0}
\end{figure}

HD41004 is a very interesting system, because the known giant planet's orbit is close enough to the HZ to significantly influence additional potentially habitable planets. First investigations concerning the stability in the region of the HZ of HD41004A were done by \citet{pilat} showing mean motion and  secular perturbations due to the interactions of the discovered GG and the secondary star. The study shows different dynamical maps for various semi-major axes of the gas planet where the appearance of mean motion and secular perturbations strongly depend on the position of the GG. In some cases these perturbations interfere with the HZ. Due to uncertainties and changes in the orbital parameters new computations of this planetary system were performed (see \citet{pilat1}) for the different planetary configurations. This study revealed the possible maximum eccentricity of the binary stars for the different orbital parameters given by the observations.\\
Even if preliminary calculations of additional hypothetical terrestrial planets in this binary system suggested a zero probability for other
planets in the HZ due to stability issues slight variations in the nominal eccentricity and semi-major axis change this picture completely (see \citet{pilat2}).\\
The nominal eccentricity of the (GG) is $e$ = 0.39 $\pm$ 0.17 the planet's orbit is rather badly constrained.
The same holds for the binary star's orbit. Therefore, we investigated whether any combination of permissible eccentricities of the system
could yield stable zones in the HZ.
Fixing the remaining orbital parameters of HD 41004 as given in Table~\ref{init}, 
we studied this system with decreased eccentricities of both the secondary ($e_{Bin}$ = 0.2) and the GG ($e_{GG}$ = 0.2).\\
For all integrations we started mass-less test planets in a region between 0.2 and 1.3 au, which covers most of the circumprimary HZ of HD 41004, see Table~\ref{hztab}. 
Since the region of the HZ is unstable when using the published orbital parameters of HD41004 we continued the computations for smaller semi-major axes in order to find the border of stability. 
We stopped our integrations at 1.3 au due to the presence of a GG at 1.64 au. 
The results are summarized in Figure~\ref{e_hd41004}, where on the $x$-axis the semi-major axis of the test planets is shown and on the $y$-axis their inclination with respect to the binaries plane, which could also be an important factor.
As one can see in the top panel of Figure \ref{e_hd41004}, upper graph no stable motion is possible in the region of the HZ for the given orbital parameters. When the secondary's eccentricity is decreased ($e_{Bin}$ = 0.2, Figure \ref{e_hd41004}, middle graph) the border of stable motion shifts just slightly outwards, but here one can also see the influence of the secondary, caused by a secular resonance between $a$ $\approx$ 0.25 and 0.5 au. This secular resonance becomes much better visible in Figure \ref{e_hd41004}, lower graph and is investigated in detail by \cite{lohinger14}. In Figure \ref{e_hd41004}, lower graph one can see clearly that the dominant influence on the HZ comes from the GG, since the stable region is shifted significantly outwards for a lower eccentricity of the GG ($e_{GG}$ = 0.2). Nevertheless the region of the HZ is still - apart from a few orbits just on the inner edge - mainly unstable.\\
Thus, we investigated the influence of changes in the semi-major axis of the known GG. Since the region of the HZ turned out to be most stable at the inner edge of the HZ we fixed in a first step the semi-major axis of the test planet at 0.8 au and varied the semi-major axis of the GG between 1.5 and 5 au. The integrations were done for 5 $\times{}$ $10^{6}$ years for $e_{Bin}$ = 0.2 and 0.4 and for $e_{GG}$ = 0.2 and 0.4. This test integrations showed, that long term stable motion in the HZ is just possible if $e_{Bin}$ = $e_{GG}$ = 0.2. For higher eccentricities no stable motion of the test planets could be found - if the GG's semi-major axis is low it perturbs the test planet and when the semi-major axis of the GG becomes higher it is perturbed itself by the secondary, which again leads to perturbations on the test planet.
For the best case scenario ($e_{Bin}$ = $e_{GG}$ = 0.2) we constructed a grid of initial conditions, where on the $x$-axis the semi-major axis of the test planets and on the $y$-axis the semi-major axis of the GG are shown. The results are displayed in Figure~\ref{atpaggi0}, where the white line indicates the current position of HD 41004 b.
Here some stable regions can be seen also for the given semi-major axis of HD 41004 A b. A detailed investigation of the dynamical stability in the system HD 41004 can be found in \cite{lohinger14}, but here we will concentrate on habitability considerations. 

\subsection{The HZ of HD 41004 A}
\label{sec:hd41004}
\begin{figure}
\includegraphics[width=7.3cm]{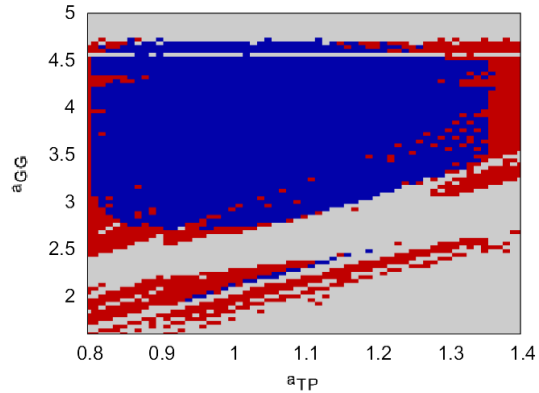}
\caption{Same as Figure~\ref{atpaggi0}, but the color code shows permanent habitability (PHZ, blue), and non-habitable regions (red).
The latter are dynamically stable but not in the PHZ. Gray areas denote dynamically unstable regions. See online version for color figures.}
\label{atpaggi0hz}
\end{figure}

\begin{figure*}
\includegraphics[width=5.0cm, angle=270]{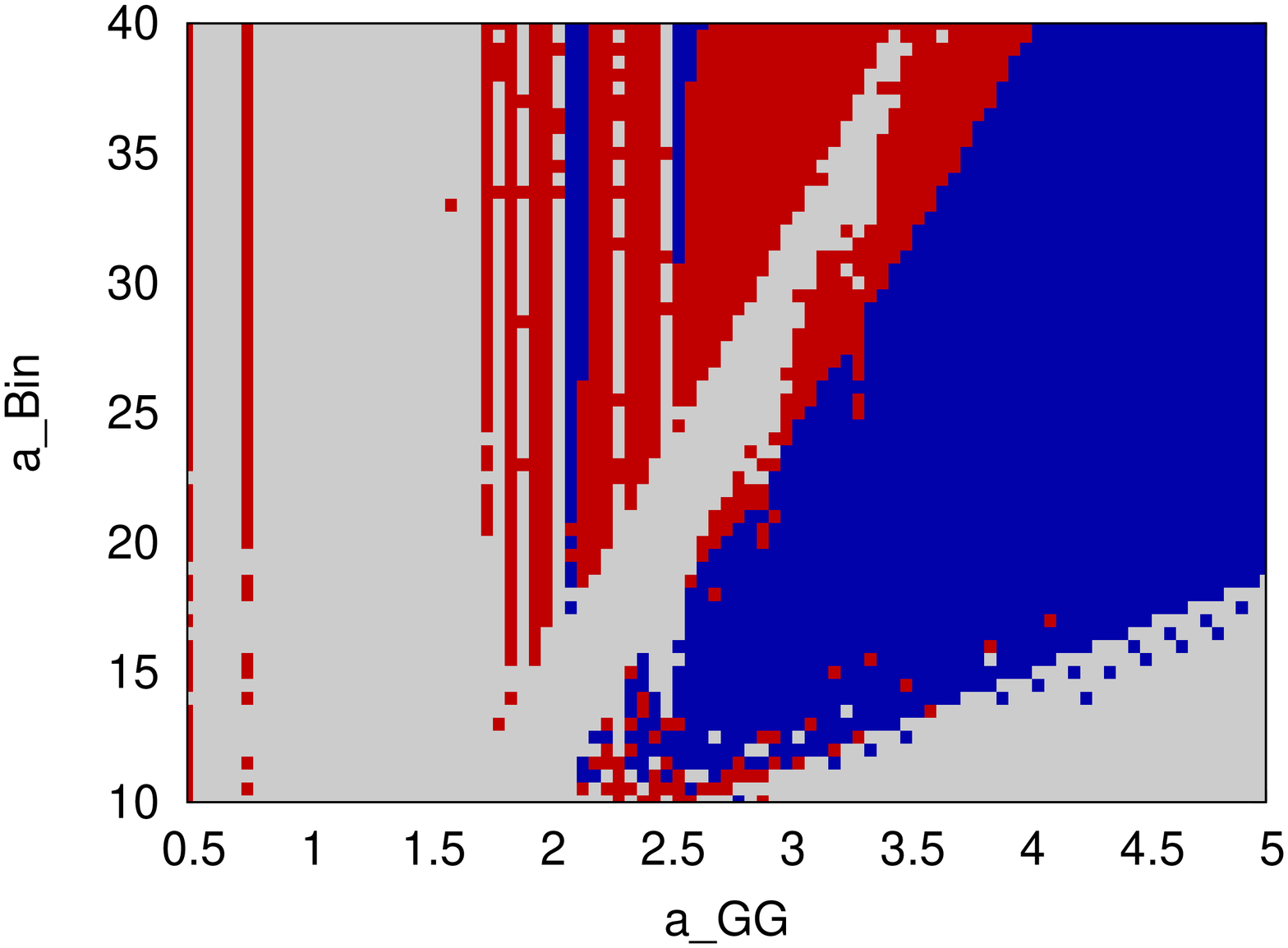}
\includegraphics[width=5.0cm, angle=270]{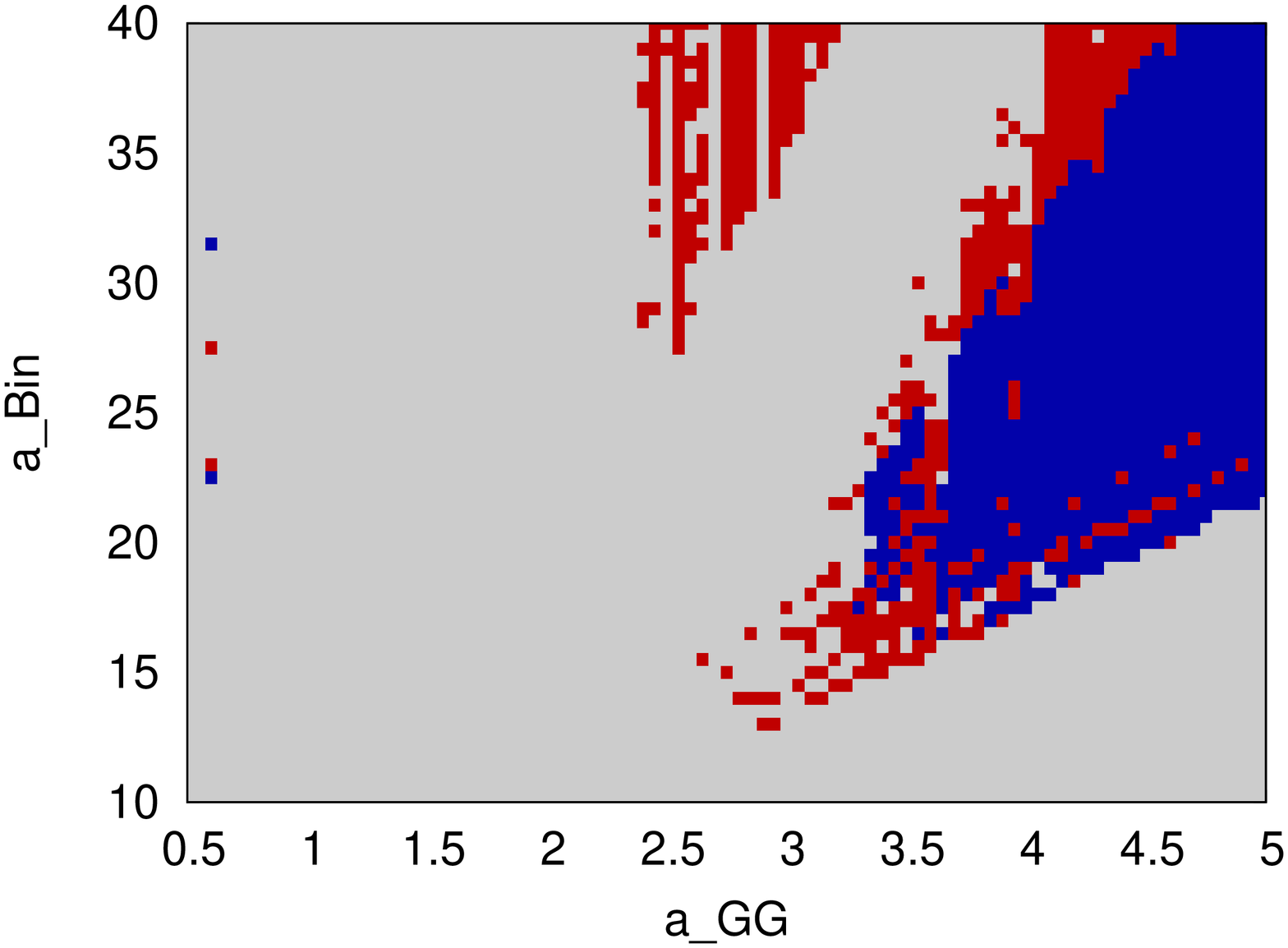}
\includegraphics[width=5.0cm, angle=270]{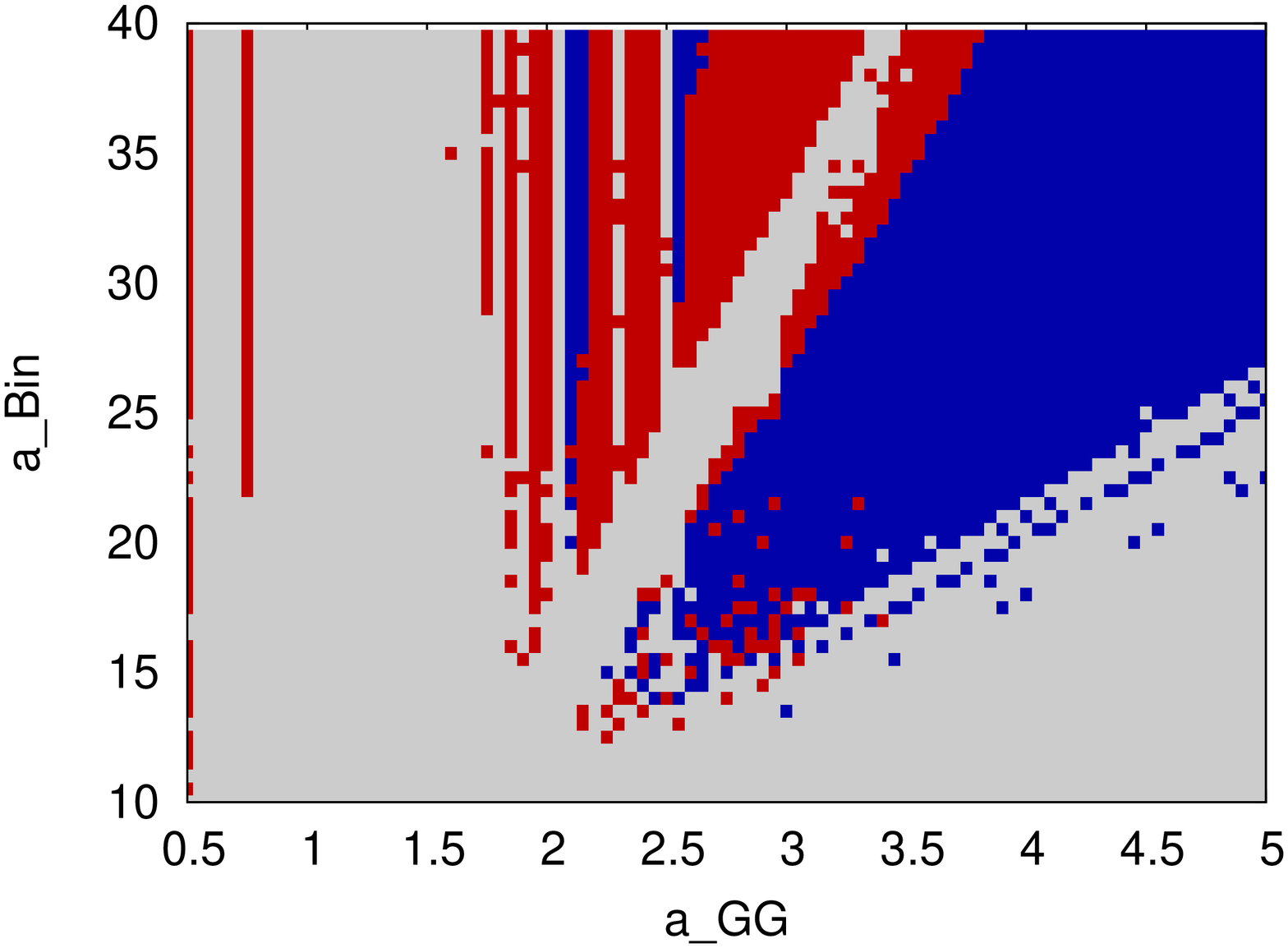}
\includegraphics[width=5.0cm, angle=270]{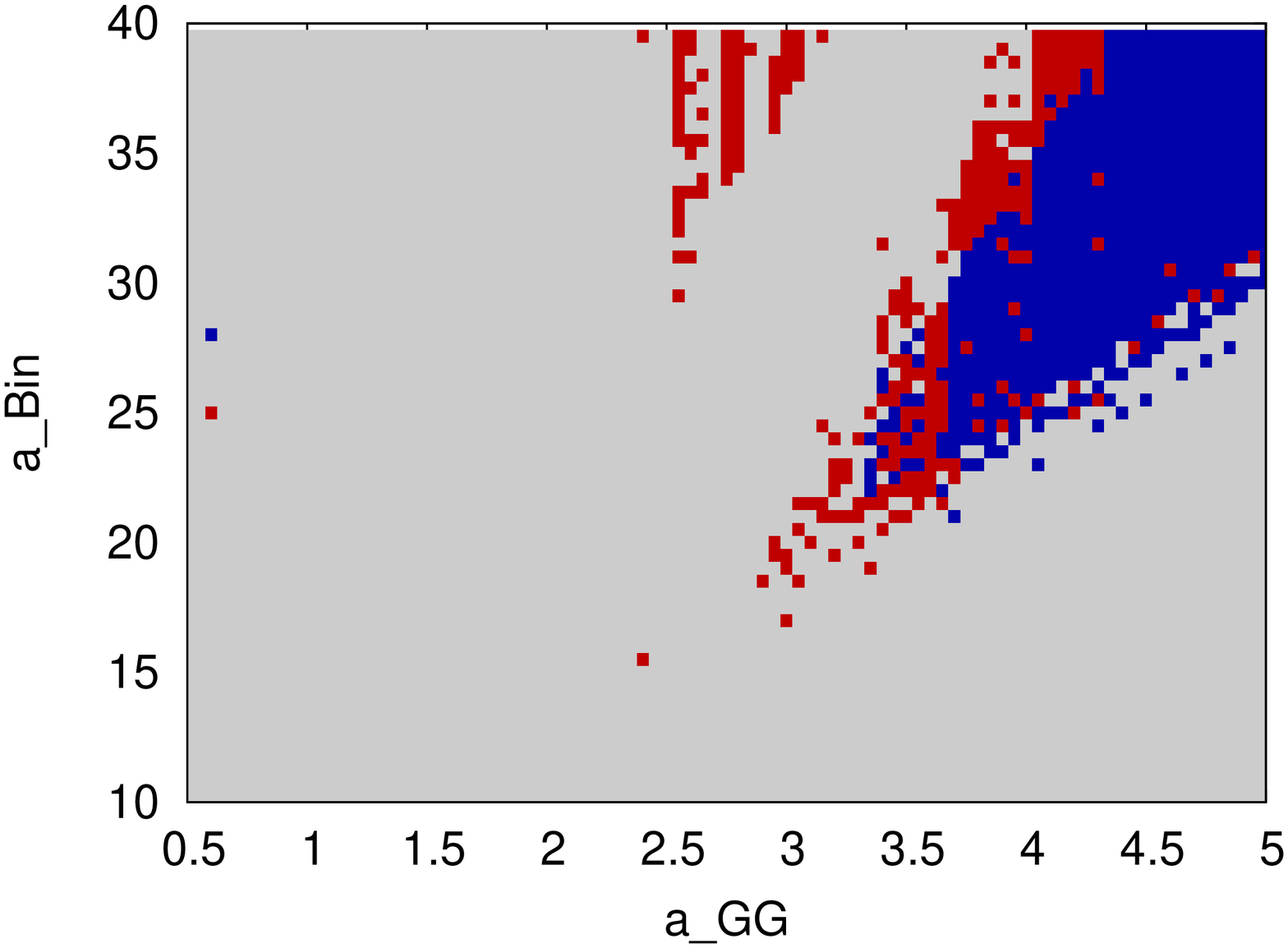}
\caption{Semi-major axis of the GG versus semi-major axis of the binary HD 41004. The left graphs show the result for an eccentricity of the GG of $e_{GG} = 0.2$ and the right graphs for $e_{GG} = 0.4$. The upper row gives the results for an eccentricity of the binary of $e_{Bin} = 0.2$ and the lower row for $e_{Bin} = 0.4$. The test-planet was always at 1.0 au; all other initial conditions were set to the values given in Table~\ref{init}. The color code shows the habitability, where blue corresponds to the PHZ, red to regions, which are dynamically stable but not in the PHZ and gray to dynamically unstable regions. See online version for color figures.}
\label{msec04}
\end{figure*}

Following the results of the previous section, we aim to ascertain that the stable regions in Figure~\ref{atpaggi0} also fulfill the criteria for habitability. 
In section \ref{hz} we defined the three body HZ including the influence of the secondary but neglecting the GG. Hence, 
we have to take the influence of the GG into account now. 
\cite{eggl} already showed that the PHZ of a specific planet in a binary system is a function of the pericenter and apocenter distances from the primary star. In hierarchical systems, those are mostly determined
by the planet's eccentricity evolution.  In our case the eccentricity of the test planet does not only change because of the influence of the secondary, but also due to the perturbations of the GG. 
Using the output of the dynamical investigations we define the region of PHZ via the maximum eccentricity values. Results are presented in Figure~\ref{atpaggi0hz}, which shows graphs similar to Figure~\ref{atpaggi0}.
From Figure~\ref{atpaggi0hz} it becomes clear that stable configurations for a GG's semi-major axis of 1.64 au are not in the PHZ. 
Nevertheless, we could find some stable, habitable configurations, e.g. when the GG is at 2 au with an eccentricity of 0.2.\\
Finally, we investigated the influence of the binaries mutual distance. To this end, we fixed the test planets position at 1 au and varied the semi-major axis of the GG and of the secondary. 
The test computations were again done for different eccentricities of the GG and the secondary ($e_{GG}$ = 0.2 and 0.4; $e_{Bin}$ = 0.2 and 0.4). 
All other initial conditions are provided in Table~\ref{init}. The corresponding results are shown in Figure~\ref{msec04}, where the color code again corresponds to the potential habitability. 
For a GG's eccentricity of 0.2 (left graphs) stable habitable configurations are possible for a GG's distance larger than $\approx$ 2 au. Interesting is that the GG's distance needs to be larger for larger distances of the secondary to allow stable configurations in the PHZ ($a_{GG} \approx$ 2 au for $a_{Bin}$ = 10 au up to $a_{GG} \approx$ 3.8 au for $a_{Bin}$ = 40 au). This is caused by the perturbing secular frequency of the GG ($g$ = $g_{GG}$, $g$ can be found in \cite{lohinger08}), which is visible as the gray diagonal region in Figure~\ref{msec04} (also visible as the arc like structure in Figure~\ref{e_hd41004} bottom graph). A detailed investigation of this phenomenon can be found in \cite{lohinger14}. However two mean motion resonances between the GG and test planet allow stable habitable configurations also within the secular resonance of the secondary, namely the 3:1 mean motion resonance at $a_{GG}$ = 2.08 au and the 4:1 mean motion resonance at 2.52 au.\\
For a GG's eccentricity of 0.4 (right graphs) stable, habitable configurations are possible for a GG's distance of more than $\approx$~3.3 au. 
Once more, the perturbing influence of the aforementioned secular resonance becomes evident.

\subsection{The HZ of HD 41004 B}
HD 41004 B is itself a binary system with a GG (HD 41004 Bb) orbiting its host star at a distance of only roughly 0.02 au, see Table~\ref{init}. 
Since the three body HZ (star star planet) for the $B$ component encompasses the region
between 0.347-0.730 au, the question arises whether HD 41004 Bb can significantly influence an Earth-like planet and change the borders of the analytically determined HZ.
To answer this question, we numerically calculated the variation in orbital elements of a massive Earth-like planet on an initially circular orbit at the inner edge of the HZ of HD 41004 B. 
Assuming that this system is coplanar, the ME of the Earth-like 
planet at the inner edge of the HZ only grows up to $e_{max}\sim0.011$, and the variations in its semimajor axis remain below $10^{-3}$ au. 
Hence, we conclude that the GG HD 41004 Bb does not influence planets in the HZ significantly and the HZ borders remain accurate. This in turn means that 
it is not impossible to have potentially habitable planets in stable orbits around HD 41004 B.

\section{Summary and Discussion}
The aim of our work was to investigate the possible existence of additional terrestrial planets in the Habitable Zones of four known close binary systems known to host giant planets. 
The selected systems were $\gamma$ Cephei, Gliese 86, HD 41004, and HD 196885. 
In a first step we calculated three body HZs for all systems excluding the influence of the known GG. 
Our investigation of circumprimary HZs showed that in two of the systems (HD 41004 and HD 196885) the GG are situated within or near the three body HZs, 
while in the third system (Gliese 86) the distance between the circumprimary HZ and the GG is large enough that the influence of the latter becomes negligible. The current orbital configuration of the $\gamma$ Cephei system does not allow for stable planetary motion in the circumprimary HZ.\\
Numerical investigations for the system Gliese 86 showed that planets in the Habitable Zone are not so much affected by the known GG. The effect of the orbital eccentricity of the secondary star is more important, but only if $e_{Bin} > 0.7$.  
Regarding the systems HD 41004 and HD 196885 we could show that the system HD 196885 does not allow any stable orbit in its HZ, even not if the eccentricities of the GG and the binary would be smaller.\\
The situation is better for the system HD 41004. Although, first integrations showed also that no stable orbits are possible in the region of the HZ, 
here small changes (within the given error bars) in eccentricity and semi-major axis of the GG and the binary could lead to at least some small stable regions in the HZ. Therefore we can conclude, that HD 41004 although not very good, is still the best candidate for stable motion in its HZ among the four known close binary systems hosting a extrasolar planet.
In all cases, HZs around the less massive component of the double star seem to be capable of hosting dynamically stable terrestrial planets.

\section*{Acknowledgments}
The authors would like to acknowledge the support from the Austrian FWF projects P22603-N16 and S11608-N16 (subproject of the NFN S116).


\begin{thebibliography}{}

\bibitem[\protect\citeauthoryear{Agol}{2011}]{agol}
Agol, E., 2011,	ApJL, 731, L31

\bibitem[{Anglada-Escud{\'e} {et~al}\mbox{.}(2012)Anglada-Escud{\'e},
  Arriagada, Vogt, Rivera, Butler, Crane, Shectman, Thompson, Minniti,
  Haghighipour, {et~al.}}]{anglada-et-al-2012}
Anglada-Escud{\'e} G. {et~al.}, 2012, The Astrophysical Journal Letters, 751,
  L16

\bibitem[{{Armstrong} {et~al}\mbox{.}(2014){Armstrong}, {Osborn}, {Brown},
  {Faedi}, {G{\'o}mez Maqueo Chew}, {Martin}, {Pollacco}, \&
  {Udry}}]{armstrong-et-al-2014}
{Armstrong} D.~J., {Osborn} H.~P., {Brown} D.~J.~A., {Faedi} F., {G{\'o}mez
  Maqueo Chew} Y., {Martin} D.~V., {Pollacco} D., {Udry} S., 2014, \mnras, 444,
  1873
  
\bibitem[\protect\citeauthoryear{Barnes \& Heller}{2013}]{barnes}
Barnes, R., Heller, R., 2013,	Astrobiology, 13, 279	

\bibitem[\protect\citeauthoryear{Beuermann et al.}{2012}]{beuermann}
Beuermann, K., Dreizler, S., Hessman, F.V., Deller, J., 2012,	A \& A, 543, A138


\bibitem[{Burke {et~al}\mbox{.}(2014)Burke, Bryson, Mullally, Rowe,
  Christiansen, Thompson, Coughlin, Haas, Batalha, Caldwell, Jenkins, Still,
  Barclay, Borucki, Chaplin, Ciardi, Clarke, Cochran, Demory, Esquerdo, Thomas
  N.~Gautier, Gilliland, Girouard, Havel, Henze, Howell, Huber, Latham, Li,
  Morehead, Morton, Pepper, Quintana, Ragozzine, Seader, Shah, Shporer,
  Tenenbaum, Twicken, \& Wolfgang}]{burke-et-al-2014}
Burke C.~J. {et~al.}, 2014, The Astrophysical Journal Supplement Series, 210,
  19
  
\bibitem[\protect\citeauthoryear{Chauvin et al.}{2006}]{chauvin06}
Chauvin, G., Lagrange, A.M., Udry, S., Fusco, T., Galland, F., Naef, D., Beuzit, J.-L., Mayor, M., 2006, A \& A, 456, 1165

\bibitem[\protect\citeauthoryear{Chauvin et al.}{2007}]{chauvin07}
Chauvin, G., Lagrange, A.M., Udry, S., Mayor, M., 2007, A \& A, 475, 723

\bibitem[\protect\citeauthoryear{Chauvin et al.}{2011}]{chauvin}
Chauvin, G., Beust, H., Lagrange, A.-M., Eggenberger, A., 2011, A \& A, 528, A8

\bibitem[\protect\citeauthoryear{Correia et al.}{2008}]{correia}
Correia, A.C.M., Udry, S., Mayor, M., Eggenberger, A., Naef, D., Beuzit, J.-L., Perrier, C., Queloz, D., Sivan, J.-P., Pepe, F., Santos, N.C., S\'egransan, D., 2008, A \& A, 479, 271

\bibitem[\protect\citeauthoryear{Cuntz}{2014}]{cuntz}
Cuntz, M., 2014, ApJ, 780, 14

\bibitem[{{Doyle} {et~al}\mbox{.}(2011){Doyle}, {Carter}, {Fabrycky},
  {Slawson}, {Howell}, {Winn}, {Orosz}, {Prsa}, {Welsh}, {Quinn}, {Latham},
  {Torres}, {Buchhave}, {Marcy}, {Fortney}, {Shporer}, {Ford}, {Lissauer},
  {Ragozzine}, {Rucker}, {Batalha}, {Jenkins}, {Borucki}, {Koch}, {Middour},
  {Hall}, {McCauliff}, {Fanelli}, {Quintana}, {Holman}, {Caldwell}, {Still},
  {Stefanik}, {Brown}, {Esquerdo}, {Tang}, {Furesz}, {Geary}, {Berlind},
  {Calkins}, {Short}, {Steffen}, {Sasselov}, {Dunham}, {Cochran}, {Boss},
  {Haas}, {Buzasi}, \& {Fischer}}]{doyle-et-al-2011}
{Doyle} L.~R. {et~al.}, 2011, Science, 333, 1602


\bibitem[\protect\citeauthoryear{Dvorak et al.}{2003}]{dvorak2}
Dvorak, R., Pilat-Lohinger, E., Funk, B, Freistetter, F., 2003, A \& A, 398, L1

\bibitem[\protect\citeauthoryear{Eggl \& Dvorak}{2010}]{34}
Eggl, S., Dvorak, R., 2010, LNP, 790, 431

\bibitem[\protect\citeauthoryear{Eggl et al.}{2012}]{eggl}
Eggl, S., Pilat-Lohinger, E., Georgakarakos, N., Gyergyovits, M., Funk, B., 2012, ApJ, 752, 11

\bibitem[{{Eggl} {et~al}\mbox{.}(2013){Eggl}, {Haghighipour}, \&
  {Pilat-Lohinger}}]{eggl-et-al-2013b}
{Eggl} S., {Haghighipour} N., {Pilat-Lohinger} E., 2013, \apj, 764, 130

\bibitem[\protect\citeauthoryear{Els et al.}{2001}]{els}
Els, S.G., Sterzik, M.F., Marchis, F., Pantin, E., Endl, M., K\"urster, M., 2001, A \& A, 370, L1

\bibitem[\protect\citeauthoryear{Endl et al.}{2011}]{endl}
Endl, M., Cochran, W.D., Hatzes, A.P., Wittenmyer, R.A., 2011, Proceedings of the International Conference. AIP Conference Proceedings, 1331, 88

\bibitem[\protect\citeauthoryear{Farihi et al.}{2013}]{farihi}
Farihi, J., Bond, H.E., Dufour, P., Haghighipour, N., Schaefer, G.H., Holberg, J.B., Barstow, M.A., Burleigh, M.R., 2013, MNRAS, 430, 652

\bibitem[\protect\citeauthoryear{Fischer et al.}{2009}]{fischer}
Fischer, D., Driscoll, P., Isaacson, H., Giguere, M., Marcy, G.W., Valenti, J., Wright, J.T., Henry, G.W., Johnson, J.A., Howard, A., Peek, K., McCarthy, C., 2009, ApJ, 703, 1545

\bibitem[\protect\citeauthoryear{Funk et al.}{2008}]{funk1}
Funk, B., Schwarz, R., Pilat-Lohinger, E., S\"uli, \'A, Dvorak, R., 2008, P\&SS, 57, 434

\bibitem[{{Ginski} {et~al}\mbox{.}(2014){Ginski}, {Mugrauer}, \&
  {Neuh{\"a}user}}]{ginski-et-al-2014}
{Ginski} C., {Mugrauer} M., {Neuh{\"a}user} R., 2014, Contributions of the
  Astronomical Observatory Skalnate Pleso, 43, 410

\bibitem[\protect\citeauthoryear{Haghighipour}{2006}]{hagh1}
Haghighipour, N., 2006, ApJ, 644, 543

\bibitem[\protect\citeauthoryear{Haghighipour \& Raymond}{2007}]{haghighipour2007}
Haghighipour, N., Raymond, S.N., 2007, ApJ, 666, 436

\bibitem[\protect\citeauthoryear{Hanslmeier \& Dvorak}{1984}]{13}
Hanslmeier, A., Dvorak, R., 1984, A \& A, 132, 203

\bibitem[\protect\citeauthoryear{Hatzes et al.}{2003}]{hatzes}
Hatzes, A.P., Cochran, W.D., Endl, M., McArthur, B., Paulson, D.B., Walker, G.A.H., Campbell, B., Yang, S., 2003, ApJ, 599, 1383

\bibitem[\protect\citeauthoryear{Holman \& Wiegert}{1999}]{holman}
Holman, M.J., Wiegert, P.A., 1999, AJ, 117, 621

\bibitem[\protect\citeauthoryear{Horner et al.}{2012}]{horner}
Horner, J., Wittenmyer, R.A., Hinse, T.C., Tinney, C.G., 2012, MNRAS, 425, 749


\bibitem[{Howard {et~al}\mbox{.}(2014)Howard, Marcy, Fischer, Isaacson,
  Muirhead, Henry, Boyajian, von Braun, Becker, Wright, \&
  Johnson}]{howard-et-al-2014}
Howard A.~W. {et~al.}, 2014, The Astrophysical Journal, 794, 51

\bibitem[{{Jaime} {et~al}\mbox{.}(2014){Jaime}, {Aguilar}, \&
  {Pichardo}}]{jaime-2014}
{Jaime} L.~G., {Aguilar} L., {Pichardo} B., 2014, \mnras, 443, 260


\bibitem[\protect\citeauthoryear{Jones et al.}{2006}]{jones}
Jones, B.W., Sleep, P.N., Underwood, D.R., 2006, ApJ, 649, 1010

\bibitem[\protect\citeauthoryear{Kaltenegger \& Sasselov}{2011}]{kaltenegger}
Kaltenegger, L., Sasselov, D., 2011, ApJ, 736, L25

\bibitem[\protect\citeauthoryear{Kaltenegger \& Haghighipour}{2013}]{kaltenegger2013}
Kaltenegger, L., Haghighipour, N., 2013, ApJ, 777, 165

\bibitem[\protect\citeauthoryear{Kasting et al.}{1993}]{kasting}
Kasting, J.F., Whitmire, D.P., Reynolds, R.T., 1993, Icarus, 101, 108

\bibitem[\protect\citeauthoryear{Kopparapu et al.}{2014}]{kopparapu}
Kopparapu, R.K., Ramirez, R.M., SchottelKotte, J., Kasting, J.F., Domagal-Goldman, S., Eymet, V., 2014, AJL, 787, L29

\bibitem[{{Kostov} {et~al}\mbox{.}(2014){Kostov}, {McCullough}, {Carter},
  {Deleuil}, {D{\'{\i}}az}, {Fabrycky}, {H{\'e}brard}, {Hinse}, {Mazeh},
  {Orosz}, {Tsvetanov}, \& {Welsh}}]{kostov-et-al-2014}
{Kostov} V.~B. {et~al.}, 2014, \apj, 784, 14


\bibitem[\protect\citeauthoryear{Lammer et al.}{2009}]{lammer}
Lammer, H., Bredehöft, J.H., Coustenis, A., Khodachenko, M.L., Kaltenegger, L., Grasset, O., Prieur, D., Raulin, F., Ehrenfreund, P., Yamauchi, M., Wahlund, J.-E., Grießmeier, J.-M., Stangl, G., Cockell, C.S., Kulikov, Yu.N., Grenfell, J.L., Rauer, H., 2009, A \& AR, 17, 181

\bibitem[{Latham {et~al}\mbox{.}(1989)Latham, Mazeh, Stefanik, Mayor, \&
  Burki}]{latham-1989}
Latham D.~W., Mazeh T., Stefanik R.~P., Mayor M., Burki G., 1989

\bibitem[\protect\citeauthoryear{Lichtenegger}{1984}]{20}
Lichtenegger, H., 1984, CMDA, 34, 357

\bibitem[\protect\citeauthoryear{Menou \& Tabachnik}{2003}]{menou}
Menou, K., Tabachnik, S., 2003, ApJ, 583, 473

\bibitem[\protect\citeauthoryear{Mugrauer \& Neuh\"auser}{2005}]{mugrauer}
Mugrauer, M., Neuh\"auser, R., 2005, MNRAS, 361, L15

\bibitem[\protect\citeauthoryear{Musielak et al.}{2005}]{musi}
Musielak, Z.E., Cuntz, M., Marshall, E.A., Stuit, T.D., 2005, A\&A, 434, 355

\bibitem[\protect\citeauthoryear{Neuh\"auser et al.}{2007}]{neuh}
Neuh\"auser, R., Mugrauer, M., Fukagawa, M., Torres, G., Schmidt, T., 2007, A \& A, 462, 777

\bibitem[\protect\citeauthoryear{Pilat-Lohinger}{2005}]{pilat}
Pilat-Lohinger, E., 2005, Proceedings of Iau Colloquium \#197, held 31 August - 4 September, 2004 in Belgrade, Serbia and Montenegro. Edited by Z. Knezevic and A. Milani. Cambridge: Cambridge University Press, 71

\bibitem[\protect\citeauthoryear{Pilat-Lohinger \& Funk}{2006}]{lohinger}
Pilat-Lohinger, E., Funk, B., 2006, Proceedings of the 4th Austrian Hungarian Workshop on celestial mechanics, Budapest, Hungary, Edited by \'A S\"uli, F. Freistetter and A. P\'al, Published by the Department of Astronomy of the E\"otv\"os University, 103

\bibitem[\protect\citeauthoryear{Pilat-Lohinger et al.}{2008}]{lohinger08}
Pilat-Lohinger, E., S\"uli, \'A.; Robutel, P., Freistetter, F., 2008, ApJ, 681, 1639

\bibitem[\protect\citeauthoryear{Pilat-Lohinger \& Funk}{2010}]{pilat1}
Pilat-Lohinger, E., Funk, B., 2010, Dynamics of Small Solar System Bodies and Exoplanets by J. Souchay and R. Dvorak (Eds.), Lecture Notes in Physics Vol. 790 Springer Berlin / Heidelberg ISSN 1616-6361; ISBN 978-3-642-04457-1, 481

\bibitem[\protect\citeauthoryear{Pilat-Lohinger et al.}{2012}]{pilat2}
Pilat-Lohinger, E., Eggl, S., Gyergyovits, M., 2012, EGU General Assembly 2012, held 22-27 April, 2012 in Vienna, Austria., 12406

\bibitem[\protect\citeauthoryear{Pilat-Lohinger et al.}{2014}]{lohinger14}
Pilat-Lohinger, E., Funk, B., Bazs\'o, \'A., Eggl, S., 2014, MNRAS, in preparation

\bibitem[\protect\citeauthoryear{Queloz et al.}{2000}]{queloz}
Queloz, D., Mayor, M., Weber, L., Bl\'echa, A., Burnet, M., Confino, B., Naef, D., Pepe, F., Santos, N., Udry, S., 2000, A \& A, 354, 99

\bibitem[\protect\citeauthoryear{Rabl \& Dvorak}{1988}]{33}
Rabl, G., Dvorak, R.: 1988, A \& A, 191, 385

\bibitem[\protect\citeauthoryear{Rafikov}{2013}]{rafikov}
Rafikov, R.R., 2013, ApJL, 765, L8

\bibitem[\protect\citeauthoryear{Raymond et al.}{2006}]{raymond}
Raymond, S.N., Mandell, A.M., Sigurdsson, S., 2006, Science, 313, 1413

\bibitem[\protect\citeauthoryear{Reffert \& Quirrenbach}{2011}]{reffert}
Reffert, S., Quirrenbach, A., 2011, A \& A, 527, A140

\bibitem[{Rein(2014)}]{rein-2014}
Rein H., 2014, Open Exoplanet Catalogue


\bibitem[\protect\citeauthoryear{Roell et al.}{2012}]{roell}
Roell, T., Neuh\"auser, R., Seifahrt, A., Mugrauer, M., 2012, A \& A, 542, A92

\bibitem[\protect\citeauthoryear{Santos et al.}{2002}]{santos02}
Santos, N.C., Mayor, M., Naef, D., Pepe, F., Queloz, D., Udry, S., Burnet, M., Clausen, J.V., Helt, B.E., Olsen, E.H., Pritchard, J.D., 2002, A \& A, 392, 215

\bibitem[\protect\citeauthoryear{Santos et al.}{2004}]{santos}
Santos, N.C., Israelian, G., Mayor, M., 2004, A \& A, 415, 1153

\bibitem[{Schneider {et~al}\mbox{.}(2011)Schneider, Dedieu, Sidaner, Savalle,
  \& Zolotukhin}]{schneider-2011}
Schneider J., Dedieu C., Sidaner P.~L., Savalle R., Zolotukhin I., 2011, arXiv
  preprint arXiv:1106.0586

\bibitem[{{Thorp} {et~al}\mbox{.}(2014){Thorp}, {Desert}, {Baranec}, {Law},
  {Johnson}, \& {Riddle}}]{thorp-et-al-2014}
{Thorp} R., {Desert} J., {Baranec} C., {Law} N.~M., {Johnson} J.~A., {Riddle}
  R.~L., 2014, in American Astronomical Society Meeting Abstracts, Vol. 223,
  American Astronomical Society Meeting Abstracts 223, p. 152.16

\bibitem[{Thorsett {et~al}\mbox{.}(1993)Thorsett, Arzoumanian, \&
  Taylor}]{thorsett-1993}
Thorsett S., Arzoumanian Z., Taylor J., 1993, The Astrophysical Journal, 412,
  L33
	
	\bibitem[\protect\citeauthoryear{Veras \& G\"ansicke}{2014}]{veras}
Veras, D., G\"ansicke, B.T., 2014, 2014arXiv1411.6012V

  \bibitem[{Wang {et~al}\mbox{.}(2014)Wang, Fischer, Xie, \&
  Ciardi}]{wang-et-al-2014}
Wang J., Fischer D.~A., Xie J.-W., Ciardi D.~R., 2014, The Astrophysical
  Journal, 791, 111
  
  
\bibitem[{{Welsh} {et~al}\mbox{.}(2012){Welsh}, {Orosz}, {Carter}, {Fabrycky},
  {Ford}, {Lissauer}, {Pr{\v s}a}, {Quinn}, {Ragozzine}, {Short}, {Torres},
  {Winn}, {Doyle}, {Barclay}, {Batalha}, {Bloemen}, {Brugamyer}, {Buchhave},
  {Caldwell}, {Caldwell}, {Christiansen}, {Ciardi}, {Cochran}, {Endl},
  {Fortney}, {Gautier}, {Gilliland}, {Haas}, {Hall}, {Holman}, {Howard},
  {Howell}, {Isaacson}, {Jenkins}, {Klaus}, {Latham}, {Li}, {Marcy}, {Mazeh},
  {Quintana}, {Robertson}, {Shporer}, {Steffen}, {Windmiller}, {Koch}, \&
  {Borucki}}]{welsh-et-al-2012}
{Welsh} W.~F. {et~al.}, 2012, \nat, 481, 475

\bibitem[{Welsh {et~al}\mbox{.}(2014)Welsh, Orosz, Short, Haghighipour,
  Buchhave, Doyle, Fabrycky, Hinse, Kane, Kostov, {et~al.}}]{welsh-et-al-2014}
Welsh W.~F. {et~al.}, 2014, arXiv preprint arXiv:1409.1605

  
\bibitem[\protect\citeauthoryear{Williams \& Pollard}{2002}]{willi}
Williams, D.M., Pollard, D., 2002, Int. J. Astrobiol., 1, 61

\bibitem[\protect\citeauthoryear{Zucker et al.}{2004}]{zucker}
Zucker, S., Mazeh, T., Santos, N.C., Udry, S., Mayor, M., 2004, A \& A, 426, 695

\end{thebibliography}

\label{lastpage}

\end{document}